\documentclass[fleqn,usenatbib]{mnras}
\usepackage{graphicx}
\usepackage{epstopdf}
\usepackage{amsmath}
\usepackage{amssymb}
\usepackage{hyperref}
\usepackage{url}
\hypersetup{draft}
\newcommand{\be}{\begin{eqnarray}}
\newcommand{\ee}{\end{eqnarray}}
\newcommand{\hMpc}{{\ifmmode{h^{-1}{\rm Mpc}}
\else{$h^{-1}$Mpc}\fi}}

\usepackage{ae,aecompl}
\usepackage[dvipsnames]{xcolor}

\newcommand{\ra}{\textit{RadioAstron}}

\title[RadioAstron observations of the quasar B0529+483]
{The high brightness temperature of B0529+483 revealed by {\em RadioAstron} and implications for interstellar scattering}

\author[Pilipenko et al.]{\parbox{\textwidth}{
S.~V.~Pilipenko$^{1}$\thanks{E-mail: spilipenko@asc.rssi.ru}, 
Y.~Y.~Kovalev$^{1,2,3}$, 
A. S. Andrianov$^1$,
U. Bach$^3$,
S.~Buttaccio$^4$, 
P.~Cassaro$^4$, 
G.~Cim\`{o}$^{5,6}$,
P.~G.~Edwards$^7$,
M.~P.~Gawro\'nski$^8$, 
L.~I.~Gurvits$^{5,9}$, 
T.~Hovatta$^{10}$ 
D.~L.~Jauncey$^{7,11}$, 
M.~D.~Johnson$^{12}$, 
Yu.~A.~Kovalev$^1$, 
A.~M.~Kutkin$^1$,
M.~M.~Lisakov$^1$, 
A.~E.~Melnikov$^{13}$, 
A. Orlati$^{14}$,
A.~G.~Rudnitskiy$^1$,
K.~V.~Sokolovsky$^{1,15,16}$, 
C. Stanghellini$^{14}$, 
P. de Vicente$^{17}$, 
P.~A.~Voitsik$^1$, 
P.~Wolak$^8$,
G.~V.~Zhekanis$^{18}$
}
\vspace{0.4cm}\\
\parbox{\textwidth}{
$^1$Astro Space Center of Lebedev Physical Institute, Russian Academy of Sciences, Profsoyuznaja 84/32, 117997 Moscow, Russia\\
$^2$Moscow Institute of Physics and Technology, Dolgoprudny, Institutsky per., 9, Moscow region, 141700, Russia\\
$^3$Max-Planck-Institute f\"ur Radioastronomie, Auf dem H\"ugel 69, D-53121 Bonn, Germany\\
$^4$INAF, Istituto di Radioastronomia, Stazione di Noto, Contrada Renna Bassa, I-96017 Noto, Italy\\
$^5$Joint Institute for VLBI ERIC, P.O. Box 2, 7990 AA Dwingeloo, The Netherlands\\
$^6$ASTRON, Netherlands Institute for Radio Astronomy, P.O. Box 2, 7990 AA Dwingeloo, The Netherlands\\
$^7$CSIRO Astronomy and Space Science, PO Box 76, Epping, NSW 1710, Australia\\
$^{8}$Centre for Astronomy, Faculty of Physics, Astronomy and Informatics, Nicolaus Copernicus University, Grudziadzka 5,\\~~~PL-87-100 Toru\'n, Poland\\
$^{9}$Department of Astrodynamics \& Space Missions, Delft University of Technology, 2629 HS Delft, Delft, The Netherlands\\
$^{10}$Tuorla Observatory, Department of Physics and Astronomy, University of Turku, V\"ais\"al\"antie 20, 21500 Kaarina, Finland\\
$^{11}$Research School of Astronomy and Astrophysics, Australian National University, Canberra ACT, 2611, Australia\\
$^{12}$Harvard-Smithsonian Center for Astrophysics, 60 Garden Street, Cambridge, MA 02138, USA\\
$^{13}$Institute of Applied Astronomy, Russian Academy of Sciences, Kutuzov embankment 10, 191187 Sankt-Peterburg, Russia\\
$^{14}$INAF, Istituto di Radio Astronomia di Bologna, Via P. Gobetti 101, I-40129 Bologna, Italy\\
$^{15}$IAASARS, National Observatory of Athens, 15236 Penteli, Greece\\
$^{16}$Sternberg Astronomical Institute, Moscow State University, Universitetsky pr. 13, Moscow 119991, Russia\\
$^{17}$Observatorio de Yebes (IGN), Apartado 148, 19180 Yebes, Spain\\
$^{18}$Special Astrophysical Observatory of RAS, Nizhnij Arkhyz, Karachai-Circassia 369167, Russia 
}
}

\date{Accepted 2017 November 17, Received 2017 November 11, in original form 2017 August 31.}

\begin{document}
\maketitle
\begin{abstract}
The high brightness temperatures, $T_\mathrm{b}\gtrsim 10^{13}$~K, detected in several active galactic nuclei by \ra\ space VLBI observations challenge theoretical limits. Refractive scattering by the interstellar medium may affect such measurements. We quantify the scattering properties and the sub-mas scale source parameters for the quasar B0529+483.
Using \ra\ correlated flux density measurements at 1.7, 4.8, and 22~GHz on projected baselines up to 240,000~km we find two characteristic angular scales in the quasar core, about $100\,\mu$as and $10\,\mu$as. Some indications of scattering substructure are found. Very high brightness temperatures, $T_\mathrm{b}\geq 10^{13}$~K, are estimated at 4.8~GHz and 22~GHz even taking into account the refractive scattering.
Our findings suggest a clear dominance of the particle energy density over the magnetic field energy density in the core of this quasar.
\end{abstract}


\section{Introduction}

Active Galactic Nuclei (AGNs) are powered by accretion onto supermassive black holes. The associated physical processes produce jets of relativistic particles which are observed at radio wavelengths to have high apparent brightness temperatures, $T_\mathrm{b}$. Theory predicts that these are limited by inverse Compton cooling to about $10^{11.5}$\,K \citep{1981ARA&A..19..373K,1994ApJ...426...51R} multiplied by the Doppler factor, $\delta$. The latter is found to have a typical value of $\delta \sim 10$ from very long baseline interferometry (VLBI) observations of superluminal motions in relativistic jets \citep{2007ApJ...658..232C,2010A&A...512A..24S,2013AJ....146..120L}. 

The brightness temperature for a source with circular Gaussian brightness distribution is measured as $T_{\rm b} = 2 {\rm ln}2 S\lambda^2 (1+z) / (\pi k_{\rm b} \theta^2)$, where $S$ is the measured flux density, $\theta$ is the angular FWHM diameter of an emitting region, $z$ is the source cosmological redshift, $\lambda$ is the observed wavelength, and $k_{\rm b}$ is the Boltzmann constant (all values in SI units). In the case of VLBI the minimum size that can be measured is limited by the angular resolution to $\theta_\mathrm{lim}\approx\lambda / B$, where $B$ is the maximum projected baseline length in the source direction. The size of the Earth limits the brightness temperature which can be measured by ground-based VLBI to about $10^{12.5}$\,K for a typical flux density of 1\,Jy. Higher values can be directly obtained only by Space VLBI \citep{2005AJ....130.2473K}, in which one of the telescopes orbits the Earth, or by indirect measurements such as those using interstellar scintillation \citep[e.g.][]{2008ApJ...689..108L}.

The Space VLBI mission \ra\ combines a 10-meter space radio telescope (SRT) on board the satellite \textit{Spektr-R}, which is in a highly elliptical orbit with an apogee of up to 370~000 km \citep{Kardashev13}, with ground-based radio telescopes (GRTs). \ra\ provides a direct way to measure brightness temperatures much higher than $10^{13}$\,K. Recently it has been shown that some quasars, indeed, have $T_\mathrm{b}>10^{13}$\,K \citep{Kovalev16,Gomez16}. This poses a challenge for the current understanding of the physics of jets. In particular, explaining of \ra\ findings may require much higher Doppler factors than is deduced from tracking superluminal components with VLBI.

However, when dealing with extremely high angular resolution observations with Space VLBI, there is another effect that can affect brightness temperature measurements. Recently it has been shown that refractive scattering by interstellar plasma may create compact features in resolved images of extended radio sources, called refractive substructure \citep{Johnson15}. This effect has been observed in Sgr A$^{*}$ with a ground-based VLBI array \citep{Gwinn14}, and in \ra\ observations of the quasar 3C273 at 18\,cm \citep{2016ApJ...820L..10J}. The primary effect of refractive substructure is to introduce a small amount of correlated flux density on long baselines, even on baselines that would have resolved-out the unscattered source \citep{Johnson15}. 

The quasar 87GB~0529+483 (J0533+4822, hereafter B0529+483) is cataloged in the 87GB survey as having a flux density of 619$\pm$70\,mJy at 4.9~GHz, with a spectral index between 80\,cm and 6\,cm of $-$0.1 (where $S \propto \nu^\alpha$) \citep{87GB,1991ApJS...75....1B}. The flat spectrum implies compact structure which, combined with the source's location at a galactic latitude $b=+8^\circ$, raises the possibility that the source may be affected by scattering. It has been observed in 29 \ra\ experiments and significant correlated flux density has been detected on projected baselines up to 19 Earth diameters. These two facts make it a suitable object to both measure its brightness temperature and examine refractive substructure effects. 
In this paper, we analyze correlated flux densities (i.e., visibility amplitudes) measured by \ra\ for B0529+483 to extract the angular sizes and brightness temperatures of its most compact features. 
The apparent brightness temperature is estimated in the source frame by taking into account the redshift of $z=1.162$ \citep{2003AJ....125..572H} but is not corrected for Doppler boosting.
We discuss two scenarios, with and without detectable substructure created by the refractive scattering in the interstellar medium of our Galaxy.

The paper is organized as follows: in Section 2, we introduce the observations used in the current study. In Section 3, we analyze the \ra\ observations assuming no scattering. In Section 4, the refractive scattering properties are analyzed. Interpretations of our findings are discussed in Section 5. We briefly summarize our results in Section 6.

\section{Observations}
\subsection{\ra\ experiments}

The observations of B0529+483 were part of the \ra\ Key Science Program AGN Survey (Kovalev et al., in prep.). The list of experiments which contain this source and were performed before 2015 January 01 is shown in Table~\ref{tab:exper}. The survey covers three bands: L at mean frequency of 1668\,MHz ($\lambda=18$\,cm), C at 4836\,MHz ($\lambda=6.2$\,cm) and K at 22236\,MHz ($\lambda=1.3$\,cm). We note that for observations obtained before December~2012 the central frequencies were 8\,MHz lower for all three bands. During the AGN Survey, \ra\ utilized an observing mode in which two bands were recorded simultaneously. In contrast, the ground-based telescopes usually observed in one band. Exceptions to this were Effelsberg and Evpatoria, which divided their observing time between the two bands. Each experiment usually lasted 40 minutes, divided into 4 scans of approximately 600\,s each. The successful fringe detections on Space-Earth baselines are clustered around the 5 months period between October 2012 and February 2013.

\begin{table}
\caption{List of RadioAstron experiments in which B0529+483 was observed. The GRTs for which a significant correlated signal with the SRT is detected are underlined. The projected baseline is shown in Earth diameters (E.D.). Telescope abbreviations: Bd\,=\,Badary 32\,m, Gb\,=\,Green Bank 100\,m, Ef\,=\,Effelsberg 100\,m, Ev\,=\,Evpatoria 70\,m, Kl\,=\,Kalyazin 64\,m, Mc\,=\,Medicina 32\,m, Nt\,=\,Noto 32\,m, Ro\,=\,Robledo 70\,m, Sv\,=\,Svetloe 32\,m, Tr\,=\,Torun 32\,m, Wb = Westerbork Synthesis Radio Telescope, Ys\,=\,Yebes 40\,m, Zc\,=\,Zelenchukskaya 32\,m.}
\begin{tabular}{lllr}
\hline
Exp.     & Observing   & Frequency, GHz, & E.D. \\
code     & epoch       & and GRTs        & \\\hline
raes03dk & 2012 Sep 29 & 4.8: \underline{Wb}, \underline{Ys}; & 7 \\
         &            &  22: Nt, Gb, Ro\\
raes03el & 2012 Oct 25 & 1.7: \underline{Zc}, Ev; 4.8: \underline{Bd} & 3.5 \\
raes03en & 2012 Oct 16 & 1.7: \underline{Zc}, \underline{Ro}; 4.8: \underline{Bd}, \underline{Ev} & 4.5 \\
raes03eo & 2012 Oct 16 & 1.7: \underline{Zc}; 4.8: \underline{Bd}, \underline{Ev} & 5.5 \\
raes03er & 2012 Oct 24 & 1.7: Ro & 3.5 \\
raes03eu & 2012 Oct 24 & 4.8: \underline{Ys} & 2.5 \\
raes03hk & 2012 Dec 09 & 4.8: Ys; 22: Gb, Nt, Ro & 16 \\
raes03hm & 2012 Dec 10 & 4.8: Bd; 22: Gb, Zc & 18 \\
raes03hv & 2012 Dec 15 & 4.8: \underline{Ys}, \underline{Mc}; 22: \underline{Gb} & 5 \\
raes03ia & 2012 Dec 16 & 4.8: \underline{Ef}, Mc; 22: Ef, Ys & 15 \\
raes03kg & 2013 Jan 25 & 1.7: Ev; 4.8: \underline{Ef} & 5 \\
raes03kj & 2013 Jan 26 & 4.8: \underline{Ef} & 11 \\
raes03kk & 2013 Jan 27 & 1.7: \underline{Gb}, Wb; 4.8: Ef & 16 \\
raes03kn & 2013 Jan 28 & 1.7: Wb; 4.8: \underline{Ef}, Mc & 19 \\
raes03kr & 2013 Feb 02 & 1.7: \underline{Wb}, \underline{Mc}; 4.8: \underline{Ys}, Tr & 3 \\
raes03ks & 2013 Feb 02 & 4.8: \underline{Tr}; 22: Ys & 2 \\
raes03kw & 2013 Feb 03 & 1.7: \underline{Tr} & 6 \\
raes03kz & 2013 Feb 03 & 1.7: \underline{Tr}, Ev, Ro & 8 \\
raes03lb & 2013 Feb 04 & 1.7: \underline{Wb}, Bd, Tr; & 12 \\
         &             & 4.8: \underline{Sv}, Zc, Ys &  \\
raks01ct & 2013 Sep 23 & 1.7: Gb, Tr; 4.8: Gb, Ys & 16 \\
raks01cx & 2013 Sep 24 & 1.7: Zc, Bd, Gb, Tr;& 19  \\
         &             & 4.8: Sv, Ef, Ys  & \\
raks01da & 2013 Sep 24 & 1.7: Bd, Tr; 4.8: Sv, Ys & 20  \\
raks01dq & 2013 Oct 01 & 1.7: Bd, Wb, Zc, \underline{Gb}; & 12 \\
         &             & 4.8: \underline{Ef}, Ys, Sv &  \\
raks01ea & 2013 Oct 03 & 1.7: Bd, Gb, Nt, Tr;   \\
         &             & 4.8: Sv, Ev, Ef, Ys & 21 \\
raks01ex & 2013 Oct 11 & 1.7: Zc, Gb, Tr; & 21  \\
         &             & 4.8: Sv, Ev, Wb  & \\
raks08cx & 2014 Oct 15 & 1.7: Kl, Bd, Tr, Ro; & 20  \\
         &             & 4.8: Sv, Zc, Nt  & \\
raks08dc & 2014 Oct 17 & 4.8: Ef, Wb, Ys, Kl, Zc, Tr; & 15  \\
         &             &  22: Gb, Ef, Ys, Tr  & \\
raks08fv & 2014 Nov 13 & 1.7: Kl, Sv, Tr; & 13  \\
         &             & 4.8: Ys, Kl, Bd  & \\
raks08gq & 2014 Nov 20 & 1.7: Gb, Kl, Sv, Zc, Tr; & 17  \\
         &             & 4.8: Ys, Kl, Nt, Tr & \\
\hline
\end{tabular}
\label{tab:exper}
\end{table}

The correlation of all these experiments was performed at the ASC \ra\ correlator \citep{RAcorr17}. Experiments raes03kg, raes03kj, raes03kk and raes03kn (Table~\ref{tab:exper}) were also correlated using an enhanced version \citep{bruni16} of the \texttt{DiFX} correlator \citep{2011PASP..123..275D}. Our comparison of the data coming from the two correlators has shown that they produce similar results.

\begin{table*}
\begin{minipage}{17cm}
\caption{Results of correlated flux density measurements.}
\begin{center}
\begin{tabular}{llrrrrl}
\hline
Experiment	& Polarization 	& Baseline & Projected & Baseline & SNR 	& Correlated \\
code		& 		& 	   & baseline,  & P.A.,    & & flux density, \\
& & &M$\lambda$& deg & & Jy\\\hline
& & & 4.8 GHz & & & \\ \hline
raes03dk & LCP & SRT--Wb & 1431.6 & 111 &    7 & 0.059 \\
raes03dk & LCP & Wb--Ys  &   24.7 &  45 &  677 & 0.575 \\
raes03dk & RCP & Wb--Ys  &   24.7 &  45 &  678 & 0.579 \\
raes03el & LCP & SRT--Bd &  661.0 &  79 &   24 & 0.312 \\
raes03en & LCP & Ev--Bd  &   63.4 &   5 &  458 & 0.830$^a$ \\
raes03en & LCP & SRT--Bd &  880.0 & 106 &   22 & 0.276 \\
raes03en & LCP & SRT--Ev &  870.5 & 102 &   24 & 0.290$^a$ \\
raes03en & RCP & Ev--Bd  &   63.4 &   5 &  798 & 0.830$^a$ \\
raes03eo & LCP & Ev--Bd  &   77.9 & 102 &  369 & 0.780$^a$ \\
raes03eo & LCP & SRT--Bd & 1228.2 & 115 &   10 & 0.151 \\
raes03eo & LCP & SRT--Ev & 1152.4 & 116 &   11 & 0.150$^a$ \\
raes03eo & RCP & Ev--Bd  &   77.9 & 102 &  638 & 0.780$^a$ \\
raes03eu & LCP & SRT--Ys &  568.4 &  34 &   29 & 0.344 \\
raes03hv & LCP & SRT--Mc & 1052.6 & 112 &   18 & 0.332 \\
raes03hv & LCP & SRT--Ys & 1050.8 & 110 &   15 & 0.179 \\
raes03hv & LCP & Ys--Mc  &   18.7 &  16 &  283 & 0.531 \\
raes03ia & LCP & Mc--Ef  &   12.2 &   1 &  498 & 1.140$^b$ \\
raes03ia & LCP & SRT--Ef & 3016.1 & 136 &   10 & 0.044 \\
raes03kg & LCP & SRT--Ef &  953.2 & 103 &   80 & 0.378 \\
raes03kj & LCP & SRT--Ef & 2306.5 & 132 &   10 & 0.050 \\
raes03kn & LCP & Ef--Mc  &   12.2 & 154 & 1762 & 0.905 \\
raes03kn & LCP & SRT--Ef & 3932.3 & 142 &    7 & 0.030 \\
raes03kn & RCP & Ef--Mc  &   12.2 & 154 & 1798 & 0.953 \\
raes03kr & LCP & SRT--Ys &  599.5 &  16 &    9 & 0.103$^c$ \\
raes03kr & LCP & Ys--Tr  &   25.4 &  97 &  129 & 0.342$^c$ \\
raes03kr & RCP & Ys--Tr  &   25.4 &  97 &   12 & 0.037$^c$ \\
raes03ks & LCP & SRT--Tr &  430.0 &  54 &   24 & 0.458 \\
raes03lb & LCP & SRT--Sv & 2579.0 & 135 &    7 & 0.080$^c$ \\
raes03lb & LCP & Sv--Zc  &   32.4 &   7 &  272 & 0.401$^c$ \\
raes03lb & RCP & Sv--Zc  &   32.4 &   7 &  258 & 0.529$^c$ \\
raks01dq & LCP & Ef--Ys  &   21.9 &  28 & 1047 & 1.280$^b$ \\
raks01dq & LCP & SRT--Ef & 2409.7 & 157 &    8 & 0.038 \\
raks01dq & LCP & Sv--Ef  &   28.1 &  31 & 2084 & 1.003 \\
raks01dq & LCP & Sv--Ys  &   50.0 &  29 &  405 & 1.260$^b$ \\
raks01dq & RCP & Ef--Ys  &   21.9 &  28 & 1043 & 0.980$^b$ \\\hline
& & & 1.7 GHz & & & \\ \hline
raes03el & RCP & SRT--Zc &  251.4 &  80 &   22 & 0.193 \\
raes03en & LCP & Ro--Zc  &   19.5 &  60 &  761 & 0.643 \\
raes03en & RCP & SRT--Zc &  297.4 & 102 &   17 & 0.125 \\
raes03kk & LCP & Gb--Wb  &   28.8 & 116 & 2406 & 0.410 \\
raes03kk & RCP & Gb--Wb  &   28.8 & 116 & 2389 & 0.437 \\
raes03kk & RCP & SRT--Gb & 1171.1 & 138 &    7 & 0.012 \\
raes03kr & LCP & Wb--Mc  &    5.6 &  30 &  401 & 0.544 \\
raes03kr & RCP & SRT--Mc &  209.1 &  15 &   11 & 0.164 \\
raes03kr & RCP & SRT--Wb &  203.0 &  15 &   43 & 0.156 \\
raes03kr & RCP & Wb--Mc  &    5.6 &  30 &  354 & 0.542 \\
raes03kw & RCP & SRT--Tr &  460.8 & 121 &   13 & 0.108 \\
raes03kz & LCP & SRT--Tr &  561.1 & 129 &    8 & 0.054 \\
raes03kz & RCP & SRT--Tr &  561.1 & 129 &    7 & 0.059 \\ \hline
\end{tabular}
\end{center}
\label{tab:flux}
$^a$The amplitude was corrected (see \S\,\ref{sss:amp}).
$^b$The data were partly flagged.
$^c$Amplitudes of ground-ground baselines are inconsistent with other data which is an indication of calibration problem at a GRT, and so the amplitude measurement is not used in the analysis.
Telescope abbreviations are given in Table~\ref{tab:exper}.
\end{minipage}
\end{table*}

\begin{table*}
\begin{minipage}{17cm}
\contcaption{Results of correlated flux density measurements.}
\begin{center}
\begin{tabular}{llrrrrl}
\hline
Experiment	& Polarization 	& Baseline & Projected & Baseline & SNR 	& Correlated \\
code		& 	         	& 	   & baseline,  & P.A.,    & & flux density, \\
& & &M$\lambda$& deg & & Jy\\\hline
& & & 1.7 GHz & & & \\ \hline
raks01dq & LCP & Bd--Wb  &   26.1 &  36 &  486 & 0.550 \\
raks01dq & LCP & Gb--Bd  &   50.1 &  71 & 1010 & 0.481 \\
raks01dq & LCP & Gb--Wb  &   32.4 &  99 & 3088 & 0.550$^b$ \\
raks01dq & LCP & Gb--Zc  &   46.5 &  96 & 1258 & 0.520 \\
raks01dq & LCP & Zc--Bd  &   20.6 &   3 &  224 & 0.645 \\
raks01dq & LCP & Zc--Wb  &   14.2 &  88 &  634 & 0.631 \\
raks01dq & RCP & Bd--Wb  &   26.1 &  36 &  474 & 0.570 \\
raks01dq & RCP & Gb--Bd  &   50.1 &  71 &  958 & 0.515 \\
raks01dq & RCP & Gb--Wb  &   32.4 &  99 & 3033 & 0.620$^b$ \\
raks01dq & RCP & Gb--Zc  &   46.5 &  96 & 1217 & 0.541 \\
raks01dq & RCP & SRT--Gb &  853.9 & 156 &    9 & 0.016 \\
raks01dq & RCP & Zc--Bd  &   20.6 &   3 &  219 & 0.642 \\
raks01dq & RCP & Zc--Wb  &   14.2 &  88 &  640 & 0.634 \\ \hline
& & & 22 GHz & & & \\ \hline
raes03dk & LCP & Ro--Nt  &  125.0 & 118 &  110 & 2.037 \\
raes03hv & LCP & SRT--Gb & 5201.5 & 108 & 13.5 & 0.139 \\ \hline
\end{tabular}
\end{center}
\end{minipage}
\end{table*}

\subsection{Measuring amplitudes}

The post-correlation data reduction, including baseline fringe fitting, bandpass and amplitude calibration, and averaging was done using the \texttt{PIMA} package \citep{Petrov11}. This software has several important features needed for Space VLBI, described below. In addition, \texttt{PIMA} is optimized for 
batch processing.

\subsubsection{Fringe fitting}
\label{sss:ff}

The fringe-fitting procedure in \texttt{PIMA} includes the search over not only the residual delay and fringe rate between radio telescopes, but also an acceleration term. This term is important for Space VLBI since it allows us to correct for imperfect knowledge of the satellite orbit (this term is not included in the FRING or KRING tasks of AIPS). During the fringe search, the two 16-MHz wide IFs were combined together to increase the sensitivity. For all successful fringe detections in the experiments that we have analyzed, the maximum deviation of the residual delay from its median value on each baseline was 3.7\,$\mu$s, the maximum delay rate corresponds to a velocity of 3.7\,cm/s, and the maximum acceleration was $5.7\times10^{-4}$\,cm/s$^2$.

The detection status of each observation was determined by the Probability of False Detection (PFD). The correspondence between signal-to-noise ratio (SNR) and PFD was found following \cite{Petrov11}. In short, we fit the theoretical probability density distribution of the false fringe amplitude in the absence of signal on the delay--fringe rate--acceleration parameter space to the full set of non-detections of the \ra\ AGN Survey. Thus, we determine the parameters of this probability density distribution for various numbers of spectral channels, integration times, and scan lengths. Knowing these parameters, we can calculate the PFD for a given value of the SNR in each experiment. 

The detection is considered significant if the SNR corresponds to PFD~$<10^{-4}$. For one experiment, raes03kn, we found a value greater than that, but lower than $10^{-3}$. As the delay of the solution was less than 0.5\,$\mu$s and the rate was less than 0.03\,cm/s, we concluded that this also was a successful detection. 

\subsubsection{Amplitude calibration}
\label{sss:amp}

Each \ra\ experiment analyzed in this paper contains only 2 or 3 baselines with very sparse ($u,v$) plane coverage, so we cannot apply standard calibration techniques utilizing closure quantities. Instead, the \textit{a priori} amplitude calibration was applied using the antenna gains and the system temperature measurements made at each antenna during the observation. For the Evpatoria radio telescope this information was unavailable for individual experiments, so we had to use the default values of system temperatures. For two experiments raes03eo and raes03en at 4.8\,GHz the signal was detected on three baselines which form an almost degenerate triangle, which allows us to correct the Evpatoria gain. The triangle has two baselines which are much longer than the third. We assume that the amplitudes on the two longest baselines are identical and that the amplitude on the baseline which does not contain Evpatoria is calculated correctly, and use this amplitude to recalculate the gain of Evpatoria station.

We also faced a problem with Yebes measurements of the system temperatures: the amplitudes in experiment raes03kr were strongly inconsistent with those from baselines with other stations in other experiments. Taking the median over many experiments for the system temperature did not solve the problem, so we had to exclude the data of this experiment at 4.8\,GHz. The same happened for the Svetloe telescope in experiment raes03lb at 4.8\,GHz.

According to \citet{2014CosRe..52..393K} the error for \ra\ System Equivalent Flux Density (SEFD) determination is 10\% while for most ground-based telescopes used in this study it is 10--20\% (e.g., \citealt{1994A&AS..103..365B})
For the baselines containing Evpatoria on which the corrections described above were applied, we conservatively increase this error to 30\%. 

\subsubsection{Time averaging}

Choosing an appropriate fringe solution interval is an important issue for measuring the correlated flux density, especially when the fringe SNR is low. If the time is too short, the correlated signal will not be detected. If it is too long, the SNR increases \citep[e.g.,][]{clark68} but the amplitude may be underestimated due to the coherence losses. 
For each experiment we trialled several fringe solution time intervals in the range 100--600~s and selected the shortest one at which the fringes were detected with required low probability of false detection.
The resulting correlated flux density values are an average of the measured quantities for a given observing experiment; they are de-biased by us following \cite{2017isra.book.....T}.




\subsubsection{Data weights}

In order to derive parameters of the source from the observed visibility amplitudes vs.\ baseline length, we attribute to each experiment a weight which is calculated from the significance of the fringe detection ($\sim$SNR) and calibration error (which we include since the data were obtained in different experiments and with different telescopes).

\begin{figure}
\includegraphics[width=\columnwidth]{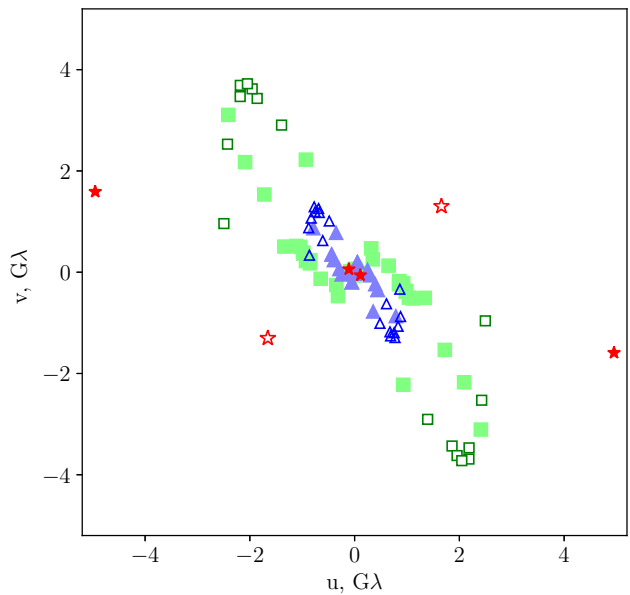}
\caption{The ($u,v$) coverage of \ra\ significant detections of B0529+483 at 1.7\,GHz (blue triangles), 4.8\,GHz (green squares), 22\,GHz (red stars). Empty symbols represent non-detections}
\label{fig:2dC}
\end{figure}

The observations are highly clustered in the $(u,v)$ plane, see Fig.~\ref{fig:2dC}. To reduce the effect of this clustering we introduce uniform weights, by computing the number of points around each point (including itself) inside a circle of radius $R$ in the ($u,v$) plane. The weight is the inverse of this number. We have tried several values of $R$ and finally adopted $R=150$\,M$\lambda$: values below this do not affect the weights significantly while values above $600$\,M$\lambda$ result in degeneracy between model parameters during approximation.

\subsection{Variability of B0529+483}

In this paper we derive the structure of the source using multi-epoch data. This is possible only if the variability of the source is weak. From Table~\ref{tab:exper} it is apparent that most successful space--ground detections cover half a year of observations. To check the variability of the source we use single dish RATAN-600 \citep[e.g.,][]{RATAN,kov99} observations, single dish data at 15\,GHz provided by the OVRO blazar monitoring program \citep{2011ApJS..194...29R}, and ground VLBI data collected by the Astrogeo Center VLBI FITS image database from \citet{2002ApJS..141...13B,2006AJ....131.1872P,2007AJ....133.1236K,2009AJ....137.3718L,2012A&A...544A..34P,2016arXiv161004951P}.

\begin{figure}
\includegraphics[width=\columnwidth]{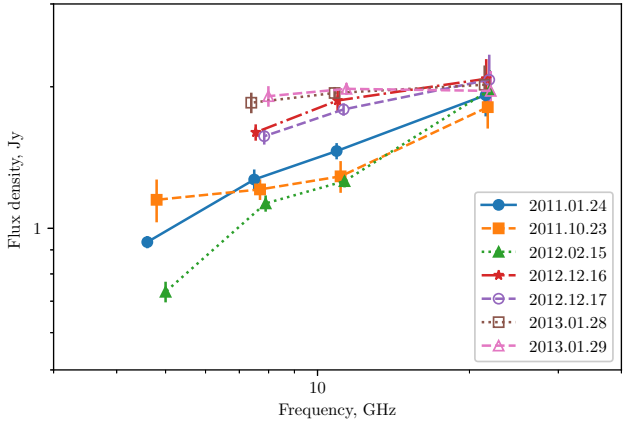}
\caption{Spectra from simultaneous RATAN-600 flux density measurements of B0529+483. Small offsets to frequencies of [4.8, 7.7, 11.1, 21.7] GHz are added to make error bars visible.}
\label{fig:ratan}
\end{figure}

The observations of RATAN-600 allowed us to obtain spectra from simultaneous flux density measurements of this quasar during the period 2011--2013 in the frequency range 4.8--22\,GHz, and four epochs of these observations lie within the period of \ra\ observations. The spectra are flat or slightly inverted, and vary with time (Fig.~\ref{fig:ratan}). Such variability indicates that it occurs in the compact core of the quasar \citep{2000PASJ...52.1027K,2002PASA...19...83K}. The flux density measured on ground--ground baselines in \ra\ experiments at 22~GHz is very close to the total flux density measured by RATAN-600 (see Table~\ref{tab:flux}). For the four epochs in 2012--2013 there are no RATAN measurements at 5~GHz, but the extrapolated RATAN-600 flux density is close to or higher than the 0.8~Jy found on ground-ground \ra\ baselines, as expected.

\begin{figure}
\includegraphics[width=\columnwidth]{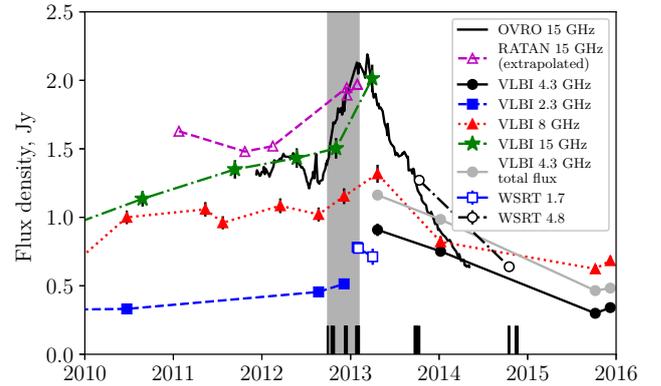}
\caption{Variability of B0529+483: 15\,GHz OVRO monitoring is shown by the solid black curve. 
     RATAN-600 data at 15\,GHz (interpolated between 11 and 22\,GHz) is shown by magenta open circles.
	 VLBI core flux densities are shown by symbols for 2.3\,GHz (filled blue squares), 4.3\,GHz (filled black circles), 8\,GHz (red triangles) and 15.4\,GHz (green stars).
	 Sum of all VLBI components at 4.8\,GHz is shown by grey circles.
	 WSRT total flux density is shown by open blue squares at 1.7\,GHz and by open black circles at 4.8\,GHz.
	 The shaded region shows the interval containing the \ra\ fringe detections. The short vertical lines at the bottom show epochs of all \ra\ observations, including non-detections.
	 } 
\label{fig:var}
\end{figure}

During 2012--2013 the flux densities measured by RATAN-600 in all bands are higher than those observed in 2011 and in the beginning of 2012. This indicates a flare in the object taking place just at the time of \ra\ observations. One can see this flare in detail in Fig.~\ref{fig:var} where we plot the RATAN-600 and OVRO data at 15\,GHz. We also plot the VLBI core flux densities extracted by fitting model components to the ground VLBI data (see section~\ref{sec:gauss}). The VLBI core flux density at 15\,GHz is found to be very close to the total flux density, indicating that most of the radiation comes from the core. A flare is observed at 15~GHz with a maximum at the beginning of 2013. At the same time, flux density variations of the VLBI core at 2--8\,GHz are less than 15 per cent, so we conclude that the variability should not strongly affect our results.

\begin{figure}
\includegraphics[width=1.01\columnwidth]{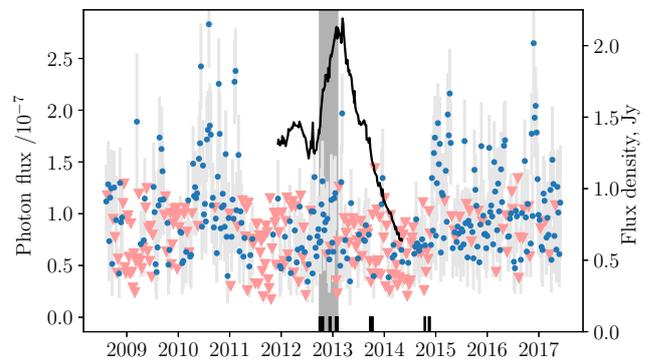}
\caption{\textit{Fermi} LAT (100\,MeV--300\,GeV) $\gamma$-ray light curve of B0529+483. The blue dots with errorbars represent weekly binned measurements, triangles -- 2-$\sigma$ upper limits. The black solid curve shows the results of the 15\,GHz radio monitoring with OVRO. The shaded region shows the interval containing the \ra\ fringe detections. The short vertical lines at the bottom show epochs of all \ra\ observations, including non-detections.}
\label{fig:gamma}
\end{figure}

The Westerbork Synthesis Radio Telescope (WSRT) participated in 8 RadioAstron VLBI experiments: three at 4.8~GHz (raes03dk, raks01ex, and raks08dc) and five at 1.7~GHz (raes03kk, raes03kn, raes03kr, raes03lb, and raks01dq). Since WSRT is able to record local array data during VLBI runs, we attempted to estimate the total flux densities of B0529$+$483. In the context of RadioAstron VLBI observations, these estimates essentially give `zero baseline' flux density values. The local array WSRT data have been processed using standard pipeline procedures\footnote{\url{http://www.astron.nl/radio-observatory/astronomers/analysis-wsrt-data/analysis-wsrt-data}}  in combination with the CASA package\footnote{\url{https://casa.nrao.edu}}. Since these WSRT observations were conducted as a piggyback on the VLBI experiments, we did not organize them with special calibration scans that involve observations of nominal amplitude calibrator sources. Instead, we relied on ad hoc use of calibration scans in adjacent observations conducted under other observing projects. In this way, we were able to process five out of the eight observations. The values of the ``zero baseline'' flux density are listed in Table~\ref{tab:wsrt}.

\begin{table}
    \caption{WSRT measured `zero baseline' flux densities}
    \begin{tabular}{llrrr}
\hline
 Exp.    & Date        &  Freq &  Peak FD & FD error (1$\sigma$) \\
 code    &             &  GHz  &  Jy      &  Jy \\
\hline
raes03kn & 2013 Jan 28 &   1.7 &  0.782  &     0.020 \\
raes03kr & 2013 Feb 02 &   1.7 &  0.774  &     0.014 \\
raes03lb & 2013 Feb 04 &   1.7 &  0.712  &     0.059 \\
\hline
raks01ex & 2013 Oct 11 &   4.8 &  1.271  &     0.030 \\
raks08dc & 2014 Oct 17 &   4.8 &  0.639  &     0.007 \\

\hline
\end{tabular}
    \label{tab:wsrt}
\end{table}

All WSRT measurements at 1.7~GHz were obtained within a week and give flux density values consistent with the other measurements described above. The two available measurements at 4.8~GHz were obtained a year apart and give significantly different (by a factor of two) values of flux density. These values and their decrease at 4.8~GHz from 2013 January to 2014 October are consistent with the RATAN and OVRO data presented in Fig.~\ref{fig:var}. Such variability on a timescale of months to years is known in B0529$+$483 and is visible in the light curves in Fig.~\ref{fig:var}. 

We also plot $\gamma$-ray data in Fig.~\ref{fig:gamma}. We used \textit{Fermi} LAT Pass~8 data to produce the 100~MeV--300~GeV weekly-binned gamma-ray light curve of B0529+483 using the ScienceTools version v10r0p5. In the event selection we followed the LAT team recommendations for Pass~8 data\footnote{\url{http://fermi.gsfc.nasa.gov/ssc/data/analysis/documentation/Pass8_usage.html}}. We modelled a $20^\circ$ region around the source using the instrument response function P8R2\_SOURCE\_V6, Galactic diffuse model ``gll\_iem\_v06.fits'', and isotropic background model ``iso\_P8R2\_SOURCE\_V6\_v06.txt''. The integral photon flux in each 7-day bin was estimated using the unbinned likelihood analysis implemented in the tool gtlike. Following \cite{abdo11}, we report a $2\sigma$ upper limit if the test statistic (TS) value \citep[e.g.,][]{mattox96} of a bin was less than 4.
All sources within $20^\circ$ of the target that are listed in the 3FGL~catalog \citep{acero15} were included in the likelihood model. We freeze the spectral parameters of all sources (including B0529+483) to the values reported in 3FGL. For sources more than $10^\circ$ from B0529+483, or sources with test statistics less than 5, also fluxes are frozen to the 3FGL values. The data in Figure 4 is weekly-binned. Very rapid flaring could be averaged out in this case, but we have also analyzed the adaptive-binned data and found no evidence of a $\gamma$-ray flare during the period of RadioAstron observations.

\subsection{Space VLBI non-detections}

From Table~\ref{tab:exper} one can see that in a number of experiments there were no detections of significant signal on SRT--ground baselines (ground stations with which there were fringe detections are underlined). In most cases when there were no detections on space--ground baselines there were significant detections on the ground--ground baselines in the same experiment. This allows us to conclude that the hardware at the ground stations was working properly. The state of the hardware on the SRT is controlled through detailed housekeeping telemetry. There were no failures detected in this way during the observations listed in Table~\ref{tab:exper}. Also, there were many experiments with significant detections of other sources adjacent in time to the non-detections for B0529+483. Thus, these non-detections are unlikely to be caused by hardware issues, so upper limits on the visibility amplitudes in these experiments can be extracted from these non-detections.

In order to determine suitable upper limits we used the measured system temperatures to calculate SEFDs of the participating telescopes. In cases when this information was not available, we used median SEFDs. Based on the results from the AGN Survey experiments (Kovalev et al., in prep.) with high signal to noise ratio, we find the coherent integration times for each band: 400\,s at 1.7 and 4.8~GHz, and 200~s at 22~GHz. At these integration times the coherence losses are less than 2 per~cent. We select the maximum $5\sigma$ upper limit for each experiment with no detection and present them in Table~\ref{tab:nodet}.

\begin{table}
\caption{Space VLBI correlated flux density upper limit estimates for experiments with no fringe detection on Space-Ground baselines. See telescope abbreviations in Table~\ref{tab:exper}.}
\begin{center}
\begin{tabular}{llrrr}
\hline
Experiment 	& Baseline & Projection	& Projection & $5\sigma$ limit, \\
code 	    & SRT-         & length, G$\lambda$& P.A., deg& mJy \\ 
\hline
 & & 4.8 GHz & & \\ 
\hline
raks01ct & Gb &  1.15 & 154 &  13 \\
raks01cx & Ef &  1.37 & 151 &  19 \\
raks01da & Ys &  2.78 & 151 &  51 \\
raks01ea & Ef &  1.51 & 149 &  19 \\
raks01ex & Ev &  4.24 & 151 &  27 \\
raks08cx & Kz &  2.75 & 148 &  49 \\
raks08dc & Ef & 14.51 & 122 &  18 \\
raks08fv & Ys &  2.61 & 111 &  36 \\
raks08gq & Ys &  1.24 & 135 &  22 \\
\hline
& & 1.7 GHz & & \\
\hline
raes03kn & Wb &  1.35 & 142 &  49 \\
raes03lb & Bd &  0.87 & 136 &  38 \\
raks01ct & Gb &  1.12 & 154 &   7 \\
raks01cx & Gb &  1.36 & 150 &   7 \\
raks01da & Tr &  1.43 & 151 &  36 \\
raks01ea & Gb &  1.50 & 149 &   7 \\
raks08cx & Kl &  1.41 & 148 &  24 \\
raks08fv & Kl &  0.93 & 111 &  24 \\
raks08gq & Gb &  1.24 & 135 &   7 \\
\hline
& & 22 GHz & & \\
\hline
raes03dk & Gb &  6.93 & 112 &  61 \\
raes03hk & Gb & 15.56 & 139 &  61 \\
raes03hm & Gb & 17.58 & 143 &  61 \\
raes03ia & Ef & 13.82 & 136 & 105 \\
raes03ks & Ys &  2.11 &  52 & 156 \\
raks08dc & Gb & 14.83 & 121 &  61 \\
\hline
\end{tabular}
\end{center}
\label{tab:nodet}
\end{table}

\section{Parsec-scale properties of B0529+483 assuming no scattering}
\subsection{Lower limits and single point measurements}

\begin{figure}
\includegraphics[trim=0cm 0.8cm 0cm 0cm,width=\columnwidth]{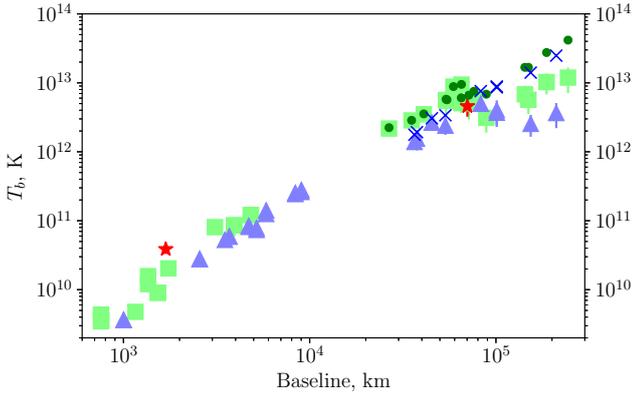}
\caption{Results of the brightness temperature measurements for B0529+483.
Lower limits in the assumption of circular Gaussian shape are calculated following \citet{Lobanov15} at 1.7\,GHz (blue triangles), 4.8\,GHz (green squares), and 22\,GHz (red stars).
The estimates of $T_\mathrm{b}$ from single-epoch experiments assuming circular Gaussian models are shown by blue crosses at 1.7\,GHz and dark-green circles at 4.8\,GHz.}
\label{fig:tbmin}
\end{figure}

\citet{Lobanov15} has shown that a minimum brightness temperature, $T_\mathrm{b,min}$, can be estimated on the basis of a single measurement of a visibility amplitude, $V_q$, at a baseline $B$, for an assumed circular Gaussian intensity distribution of the source.
These brightness temperature limits for B0529+483 are shown in Fig.~\ref{fig:tbmin}. The space-ground part of this plot covers baselines $B>10^4$~km. 
The highest lower limits are found to be $5\times10^{12}$\,K at 1.7~GHz, $1.2\times10^{13}$~K at 4.8~GHz, and $5\times10^{12}$~K at 22~GHz.
For several examples presented in \citet{Lobanov15}, the difference between the model brightness temperature and a lower limit for a resolved component is within a factor of 2.

The non-detections presented in Table~\ref{tab:nodet} cover longer time interval than our successful fringe detections. The lower limit of the brightness temperature derived from the non-detections covers the range from 
$3.5\times10^{12}$\,K to $2.3\times10^{13}$\,K at 1.7\,GHz,
from $1.8\times10^{12}$\,K to $1.9\times10^{13}$\,K at 4.8\,GHz and
from $0.8\times10^{12}$\,K to $2.4\times10^{13}$\,K at 22\,GHz. This may reflect the variability of B0529+483.

One can also estimate the brightness temperature of the compact feature of the source assuming a single circular Gaussian model for the most compact component by taking the median value of ground-ground visibility amplitudes as the amplitude of the Gaussian and determining its width from a single visibility measurement at space-ground baselines. The results of such estimates are shown in Fig.~\ref{fig:tbmin} as well with brightness temperatures rising to $3\times10^{13}$\,K at 1.7\,GHz and $5\times10^{13}$\,K at 4.8\,GHz. 

Note that both these methods can overestimate the brightness temperature of the source if there is refractive substructure from scattering \citep{Johnson15,2016ApJ...820L..10J}.

\subsection{Gaussian models}
\label{sec:gauss}

In this Section we build a more complex model which is aimed to fit all the visibility data. We use the weighted least squares method to determine model parameters which are presented in Table~\ref{tab:fit}.

\begin{figure}
\includegraphics[width=\columnwidth]{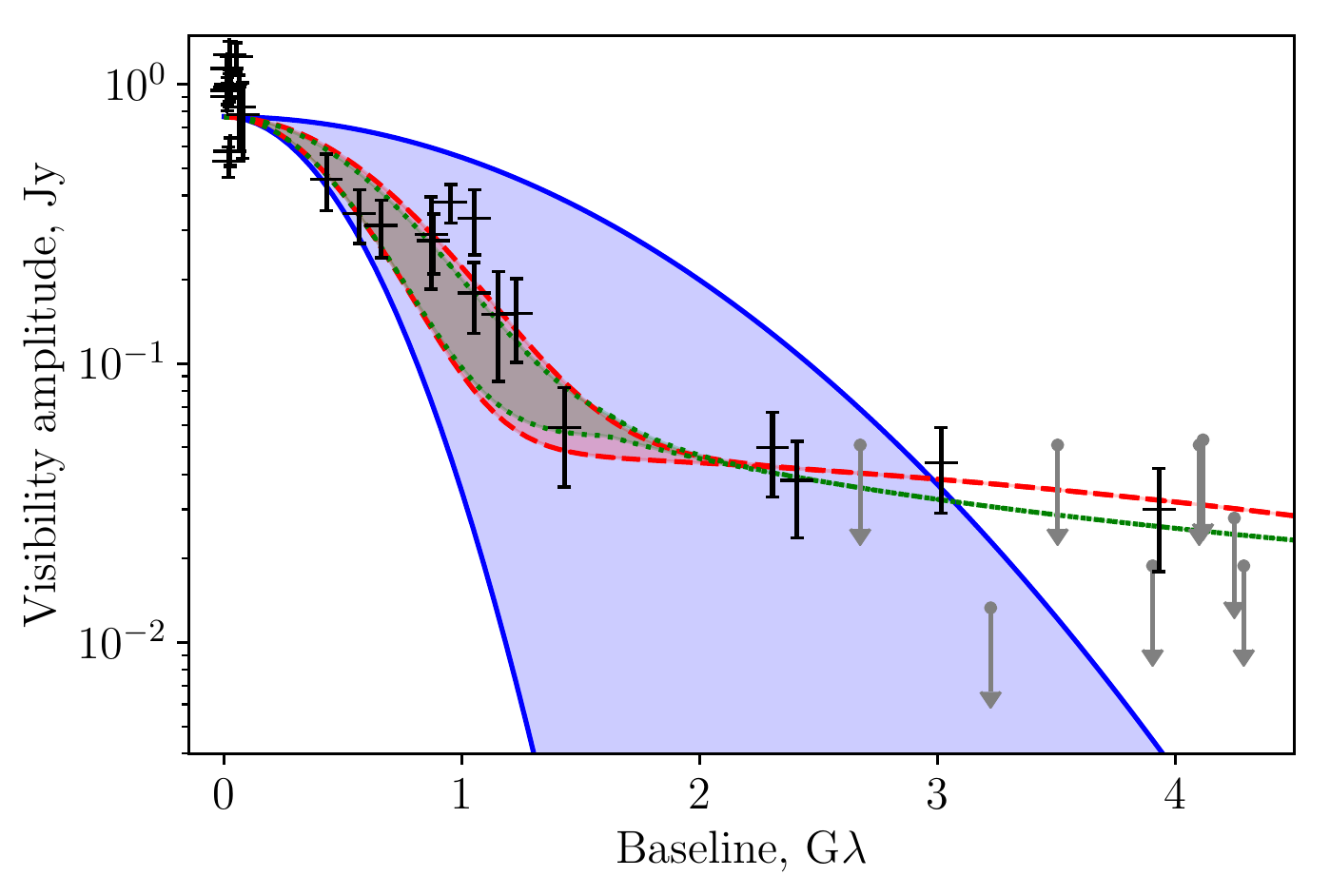}
\caption{The visibility amplitude as a function of baseline length at 4.8~GHz. Error bars represent \ra\ data, blue shaded region between solid lines -- Single elliptical Gaussian model, red region between dashed lines -- Double Gaussian model, green region between dotted lines --  model with refractive substructure. Borders of the shaded regions correspond to minor and major axes of the model, the regions itself cover visibility amplitude values for various position angles. Upper limits shown by dots with errors represent results from Table~\ref{tab:nodet}.
\label{fig:1dC}
}
\end{figure}

\begin{figure}
\includegraphics[width=\columnwidth]{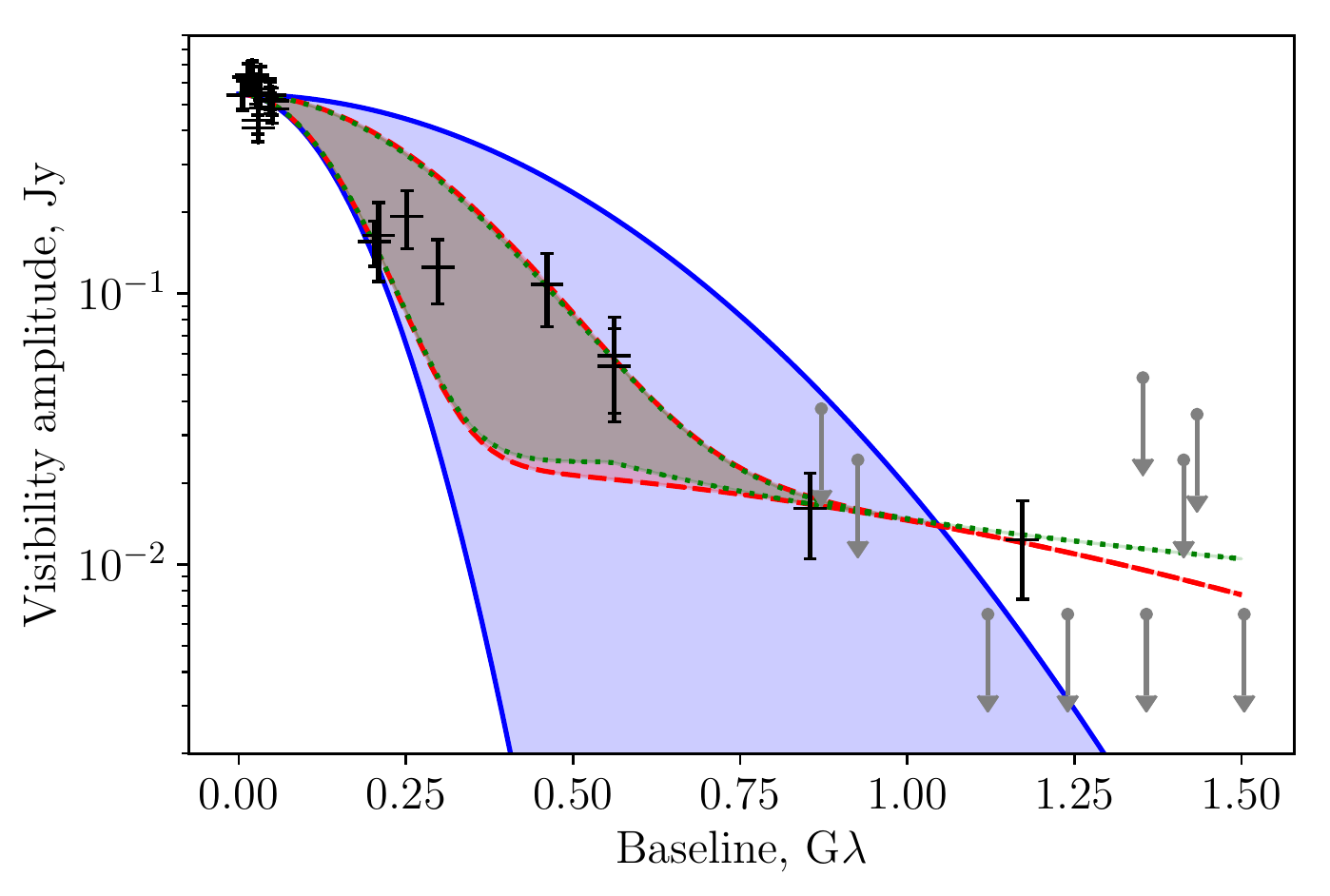}
\caption{Same as Fig.~\ref{fig:1dC} but at 1.7\,GHz.}
\label{fig:1dL}
\end{figure}

First, we start with a single elliptical Gaussian model, which is called `Single Gaussian' in Table~\ref{tab:fit}. It has quite a large value of $\chi^2_\mathrm{reduced}$ which we interpret as a result of the simplicity of the model: from Figs.~\ref{fig:1dC} and \ref{fig:1dL} one can see that the points at the longest baselines are poorly described by this model.

We add a second circular Gaussian component to this model and call it `Double Gaussian'. In this case the $\chi^2_\mathrm{reduced}$ is much more reasonable; the difference from 1.0 at 4.8\,GHz comes from a feature at $\sim 1$\,G$\lambda$ in Fig.~\ref{fig:1dC} that is not well fitted by the model. This feature is represented by only two points from two experiments (raes03hv, raes03kg). The deviations of the data points from the model are 2.6 and 1.2\,$\sigma$. This may be the result of the source variability, since the two experiments precede, but are close to the maximum of the 15\,GHz total flux density variations in Fig.~\ref{fig:var}.

From Fig.~\ref{fig:2dC} one can see that the distribution of our measurements on the ($u$,$v$) plane is highly elongated (for both detections and non-detections). From Table~\ref{tab:fit} it is seen that the minor semi-axes at the real plane of the larger components in our models are roughly coincident with the direction of this elongation. This could be due to chance. First, the orientation of elliptical components is determined by data points on baselines $<0.5$\,G$\lambda$ at 1.7\,GHz and $<1.5$\,G$\lambda$ at 4.8\,GHz. At these scales in Fig.~\ref{fig:2dC} the data points are distributed more uniformly over the position angles. Second, the orientation of our model components coincide with the orientation of the core on the ground VLBI maps, so we believe that the position angles given in Table~\ref{tab:fit} are not artifacts of the modelling.

Tables~\ref{tab:exper} and \ref{tab:flux} list several experiments at 4.8~GHz in which the signal was detected on three baselines forming a triangle (raes03en, raes03eo, raes03hv). This enables us to analyze closure phases which in principle carry significant information on the brightness distribution in the source and at the same time, under modest assumptions, are free of systematic phase noise \citep{2017isra.book.....T}. We note however that a poor ($u$,$v$)-coverage and a high degeneracy of the closure triangles (the ground--SRT baselines are almost equal and are much longer than the ground--ground baseline) significantly compromise the efficiency of the method \citep[e.g.,][]{1986AJ.....92..213L,2008A&A...480..289M}.

Nevertheless, we measured the closure phases and used them in the fitting procedure. The signal to noise ratio in the experiments discussed is low, and only in one of these experiments, raes03en, is the closure phase found to be above the detection threshold. We used the closure-phase values from all three experiments to get a consistency check of our brightness distribution model based on the analysis of the visibility amplitudes presented above. Specifically, we varied the separation between the components in the `Double Gaussian' model. Qualitatively, the closure phase models are consistent with the amplitude-based modeling. Due to the lack of phase data this result is strongly model dependent and the measured closure phase may be explained without introducing the offset between the components by, e.g., adding a small asymmetry to the largest component of the model. We note that this result should be treated with caution due to the scarcity of the available data and low SNR values. In particular, while qualitatively the closure-phase-model does seem to permit amplitude-based solutions, the two outlier points at around $\sim 1$~G$\lambda$ in Fig.~\ref{fig:1dC} are poorly fit.



The presence of signal at long space-ground baselines can also be explained by the effect of substructure created by refractive scattering on the interstellar medium inhomogeneities \citep{Johnson15}. We discuss this possibility in Section \ref{sec:ref}. Alternatively, it can arise due to the internal small-scale structure of the source \citep{Kovalev16,Gomez16}. If the detections on the longest baselines reflect intrinsic core parameters, the brightness temperature of the core is $0.4\times10^{13}$\,K at 1.7\,GHz, $1.7\times10^{13}$\,K at 4.8\,GHz and $1\times10^{13}$\,K at 22\,GHz.

The upper limits and detections at long spacings presented in Table~\ref{tab:nodet} and Figs.~\ref{fig:1dC} and \ref{fig:1dL} show significant scatter. From Fig.~\ref{fig:2dC} one can see that the baseline projection vectors of non-detections are close to that of the detections, so the scatter cannot be explained by the source assymetry. It can be explained by variations of the core flux density or its structure: the non-detections cover two years of observations while successful detections only a period of 5 months. While variation of the source's apparent angular size due to physical variation of the linear size of the source cannot be ruled out, its amplitude is limited by the causality principle (the linear size variation cannot exceed distance covered by light over the characteristic variability time). In the following analysis we do not attribute non-detections to the structural variations (i.e. increase of the compact component linear size resulting in resolving it out). We might also underestimate the detection sensitivity in some cases especially if weather conditions were poor. Due to this large scatter we do not use the upper limits to further constrain the models presented here.

In Table~\ref{tab:nodet} there is one point that deserves special discussion: in the experiment raes03ks there were no detections at 22\,GHz on relatively small projection baselines, 2.1\,G$\lambda$. In this experiment Yebes was the only ground radio telescope, so we cannot be sure that this non-detection is not caused by an unidentified hardware failure. On the other hand, on the same day Yebes gave fringes at 4.8\,GHz in another experiment. If we consider the 22\,GHz non-detection as real, it is not consistent with the circular Gaussian model for 22\,GHz data presented in Table~\ref{tab:fit}. The non-detection is located almost exactly on the largest semi-axis in the image plane of the components of `Double Gaussian' model at 4.8 and 1.6\,GHz in the Table~\ref{tab:fit} (see also empty red stars in Fig.~\ref{fig:2dC}). If we fit an elliptical Gaussian model to the 22\,GHz data taking into account this non-detection, we obtain FWHM sizes of $0.08 \times 0.03$ mas and $T_{\rm b}=0.5\times 10^{13}$. The difference from the values in Table~\ref{tab:fit} does not affect our results, so we further do not consider this updated model which includes the (uncertain) non-detection.

\begin{table*}
\begin{minipage}{16cm}
\caption{The fit parameters for the measured amplitudes. See description of the models in text. The errors are one standard deviation. The angular sizes, $\theta_a$ and $\theta_b$, are FWHMs. Position Angles are given for the largest semi-axis at the real plane. When an error is not given this means that the parameter was fixed.}
\begin{tabular}{lllllllr} \hline
Model                     & $S_0$,   & $\theta_a$, & $\theta_b$, & P.A., & $C$, & $T_{\rm b}$, & $\chi^2_\mathrm{reduced}$\\
                          & mJy          & mas     & mas         & deg   & mJy  & $10^{13}$\,K & \\ \hline
4.8 GHz & & & & & & & \\
Single Gaussian           & $770 \pm 30$ & $0.19 \pm 0.01$  & $0.06 \pm 0.001$ & $36 \pm 1$  & -- & $0.71 \pm 0.08$& 6.0 \\
Double Gaussian, comp.\ 1 & $720 \pm 40$ & $0.18 \pm 0.05$  & $0.13 \pm 0.038$ & $47 \pm 38$ & -- & $0.34 \pm 0.20$& 3.2 \\
Double Gaussian, comp.\ 2 & $ 48 \pm  6$ & $0.018\pm 0.015$ & $0.018\pm 0.015$ & --          & -- & $1.7\pm0.8$ & \\
Gaussian + Substructure   & $700 \pm 30$ & $0.19 \pm 0.05$  & $0.14 \pm 0.040$ & $51 \pm 57$ & $60 \pm 11$ & $0.30 \pm 0.18$& 3.1 \\
\hline
1.7 GHz & & & & & & & \\
Single Gaussian           & $550 \pm 20$ & $0.64 \pm 0.05$ & $0.20 \pm 0.005$ & $37 \pm 1$ & -- & $0.41 \pm 0.05$& 4.5 \\
Double Gaussian, comp.\ 1 & $520 \pm 20$ & $0.64 \pm 0.09$ & $0.31 \pm 0.028$ & $47 \pm 5$ & -- & $0.25 \pm 0.06$& 1.2 \\
Double Gaussian, comp.\ 2 & $ 24 \pm  4$ & $0.08\pm0.06$   & $0.08 \pm 0.06$  & --         & -- & $0.4\pm0.3$ \\
Gaussian + Substructure   & $520 \pm 20$ & $0.64 \pm 0.09$ & $0.32 \pm 0.027$ & $47 \pm 5$ & $23 \pm 6$ & $0.24 \pm 0.06$& 1.2 \\
\hline
22 GHz & & & & & & & \\
Gaussian (circular)       & 2200              & 0.034 & 0.034 & -- & -- & 1.0 & -- \\
\hline
\end{tabular}
\label{tab:fit}
\end{minipage}
\end{table*}

\section{Refractive scattering manifestations}
\label{sec:ref}

The main properties of interstellar scattering are summarized in reviews of \citet{1990ARA&A..28..561R} and \citet{1992RSPTA.341..151N}. We expect that there are three regimes: at highest frequencies there is no scattering, i.e. the angular broadening scale is much smaller than the angular size of the observed components of the core. At lower frequencies the weak scattering (or refractive scattering) creates an angular broadened image. In this regime the angular broadening size is expected to scale with frequency close to $\nu^{-2}$. In contrast, the \citet{BK79} model suggests a $\nu^{-1}$ scaling of the intrinsic core size. In this regime the angular broadened image should show the refractive substructure. Finally, at even lower frequencies in the regime of strong scattering the source should show rapid and strong intra-day variability, which is observed in some quasars (see, e.g., \citet{1997ApJ...490L...9K}).

Refraction introduces substructure \citep[e.g.,][]{Gwinn14,2016ApJ...820L..10J} on angular scales comparable to, and smaller than, the angular broadening scale. As shown by \citet{Johnson15}, the ratio of the correlated flux density introduced by substructure to the total flux density depends on the observed source size, the angular broadening scale, and the interferometric baseline. Refractive substructure is a stochastic effect, so the correlated flux density on long baselines from refractive substructure varies from baseline to baseline and from epoch to epoch. The amplitudes on these baselines are drawn from a Rayleigh distribution \citep{Johnson15}, and its dispersion depends on properties of the source and scattering medium.

In this section we utilize both the scaling of angular broadening with frequency and the relation between the angular broadening and the refractive substructure to constrain the scattering parameters. We give results only at 1.7 and 4.8\,GHz, since the 22\,GHz image is not affected by scattering at the angular scales of interest, as follows from our analysis in this Section. 

\subsection{Determination of the scattering parameters}

\begin{figure}
\includegraphics[width=\columnwidth]{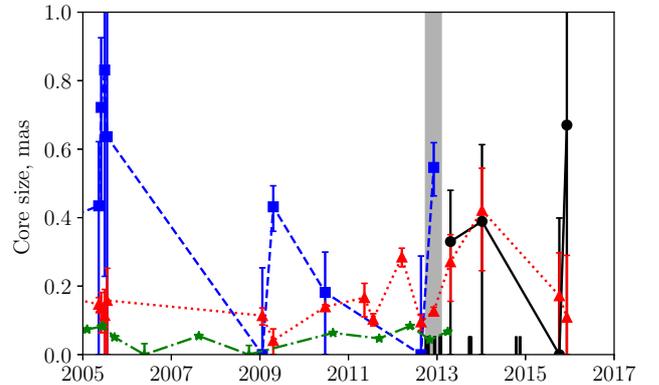}
\caption{Core size of B0529+483 as revealed by Earth-based VLBI experiments at 2.3\,GHz (squares), 4.3\,GHz (circles), 8\,GHz (triangles) and 15.4\,GHz (stars). The shaded region shows the interval containing the \ra\ fringe detections. The short vertical lines at the bottom show epochs of all \ra\ observations, including non-detections.}
\label{fig:core}
\end{figure}

\begin{figure}
\includegraphics[width=\columnwidth]{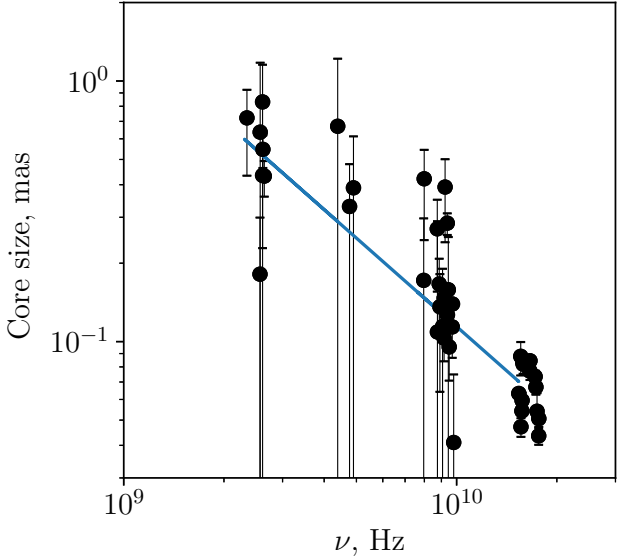}
\caption{Core size of B0529+483 vs.\ observation frequency from Earth-based experiments. A $\nu^{-1.1}$ fit is shown by the solid blue line. Small random offsets are added to frequency to make the error bars visible.}
\label{fig:core_freq}
\end{figure}

The first hypothesis we check is that the two larger components of `Double Gaussian' model at 1.7 and 4.8\,GHz in Table~\ref{tab:fit} are both angular broadened images of a more compact core. This assumption is not supported by the scaling of size of these components along the minor axis with frequency as $\nu^{-0.9}$. We also use the Earth-based VLBI data to measure core size variations over a much longer period. Earth-based VLBI has sufficient resolution to detect the first component of our `Double Gaussian' model. The results are shown in Figs.~\ref{fig:core} and \ref{fig:core_freq}. The errors at each particular epoch are quite large, but the median scaling shows a power law index of $-1.1$, which is consistent with an intrinsic core size variations, but inconsistent with scattering angular broadening. The core sizes found by Earth-based VLBI at 2 and 4\,GHz are close to those found by \ra\ at 1.7 and 4.8\,GHz, respectively, but the Earth-based data have much larger error bars.

Radio interferometer resolution also scales as $\nu^{-1}$, and to ensure that this does not affects the scaling of source components, we determine the component size at 4.8~GHz with the resolution reduced to that at 1.7~GHz, i.e. we ignore data points at $B>1.5$~G$\lambda$. This changes the component size by only 2.5~per~cent with respect to the values given in Table~\ref{tab:fit}, so we conclude that the size of resolved (the largest) component is not attenuated by the resolution effect.

A second hypothesis is that the large (first) component of the `Double Gaussian' model is an only-partially-broadened image of the core of B0529+483, while the detections on the longest baselines represent the scattering substructure. We call this hypothesis `Strong angular broadening'.

We use the properties of the detections on the longest baselines to infer the angular broadening scale, $\theta_\mathrm{scatt}$, from the properties of scattering substructure. 
If the measured flux densities at baselines $B>10^5$~km are due to refractive scattering, their values should differ from epoch to epoch and from baseline to baseline. Their amplitudes should follow Rayleigh distribution with the dispersion $\sigma \propto \sigma_\mathrm{ref} S_0 B^{-5/6}$, where $S_0$ is the zero-spacing amplitude, $B$ is the baseline projection, $\sigma_\mathrm{ref}$ is a constant that characterizes the scattering screen. In order to measure this constant, we add the following term in the models:

\be
S(u,v) = \left \{ 
\begin{tabular}{ll}
$C\times \left( {B / 10^5 \mathrm{km}} \right)^{-5/6}$, & $B>10^5$ km \\
$C$, & $B<10^5$ km,
\end{tabular}
\right.
\label{eq:term}
\ee
where $C$ is a constant. The behavior of these additional flux density variations should be more complex at baselines $B<10^5$\,km, but we ignore this in our simple model. By fitting the data to a model which consists from a single elliptical Gaussian component and a term from equation~\ref{eq:term} we determine an average $C$.
This model is denoted as `Gaussian + Substructure' in Table~\ref{tab:fit}. It is seen from the $\chi^2$ column that it describes the data much better than the single Gaussian and equally as well as the `Double Gaussian' model. The fit to this model is also shown in Figures~\ref{fig:1dC} and \ref{fig:1dL}.

We cannot directly check the statement that the amplitudes on long baselines correspond to the Rayleigh distribution due to the small number of measurements, but we can try to improve our fit by taking this fact into account. We do this by adding errors of 52 per cent (the dispersion of the Rayleigh distribution) to data points at $B>10^5$\,km. However, from the comparison of $\chi^2$ with the previous model there is little improvement in both bands.

The RMS of the refractive noise, $\sigma_\mathrm{ref}$, defined in \citet{Johnson15} is related to the measured constant C:
\be
\sigma_\mathrm{ref}={2\over \sqrt{\pi} S_0}C\,.
\label{eq:sigma-C}
\ee
Since the number of data points with $B>10^5$\,km is small (4 at 4.8\,GHz, 2 at 1.7\,GHz), we use Monte-Carlo (MC) modeling instead of equation \ref{eq:sigma-C} to relate the measured value of $C$ with the $\sigma_\mathrm{ref}$. We draw random amplitudes of points at given baseline projections from Rayleigh distribution with unit dispersion and fit them by a constant to connect the value of $C$ and its scatter to $\sigma_\mathrm{ref}$. Then, using equation (19) from \citet{Johnson15} we connect $\sigma_\mathrm{ref}$ with $\theta_\mathrm{scatt}$, the scattered size of a point source, which is used in this paper to characterize the scattering:

\be
\begin{tabular}{lll}
$\sigma_\mathrm{ref}$ & $=$ &
$0.0038 \times
\left ( {\lambda \over \mathrm{6\,cm}} \right )
\left ( {B \over 10^5\,\mathrm{km}} \right )^{-5/6}
\left ( {\theta_\mathrm{scatt} \over 30\,\mu\mathrm{as}} \right )^{5/6} \times $ \\
& & $\times \left ( {\theta_\mathrm{img} \over 300\,\mu\mathrm{as}} \right )^{-2}
\left ( {D \over 1\,\mathrm{kpc}} \right )^{-1/6}$ \\
& $=$ & $0.0071 \times
\left ( {\lambda \over \mathrm{18\,cm}} \right )
\left ( {B \over 10^5\,\mathrm{km}} \right )^{-5/6}
\left ( {\theta_\mathrm{scatt} \over 300\,\mu\mathrm{as}} \right )^{5/6} \times$ \\
& & $\times \left ( {\theta_\mathrm{img} \over 1000\,\mu\mathrm{as}} \right )^{-2}
\left ( {D \over 1\,\mathrm{kpc}} \right )^{-1/6}$, \\
\end{tabular}
\label{eq19}
\ee
where $\theta_\mathrm{img}$ is the observed image size, $D$ is the distance of the screen from the observer (1\,kpc is a typical value).
The results are shown in Table~\ref{tab:tscat}.

\begin{table}
\caption{Angular broadening of B0529+483}
\begin{center}
\begin{tabular}{lll}
\hline
Angular broadening	    &  4.8 GHz	     & 1.7 GHz 	\\ \hline
Strong		  	        & $0.20\pm 0.08$ & $0.21\pm 0.09$\\
Weak		 	        & $0.04$	     & $0.3$		\\
NE2001 \citep{Cordes02}	& 0.11		     & 1.25		\\ \hline
\end{tabular}
\end{center}
\label{tab:tscat}
\end{table}

In the `Strong angular broadening' hypothesis the scattering scale is close to the observed image size which means that this size is determined by the scattering, not by the intrinsic source parameters.
The observed angular broadening does not scale as with frequency as $\nu^{-2}$, so we conclude that this hypothesis is ruled out. 


We consider another possibility: that the angular broadening at 1.7~GHz is about 0.3~mas and this size scales to about 0.04~mas at 4.8~GHz. This size is marginally consistent with the size of the second component of `Double Gaussian' model. In this case only at 1.7~GHz do we see the refractive substructure and the angular broadening at 4.8~GHz is much smaller than in the previous case, and so we call this hypothesis `Weak angular broadening'.

In Table \ref{tab:tscat} we also put the angular broadening predictions of the \citet{Cordes02} NE2001 model of the Galactic free electron distribution. NE2001 predictions are inconsistent with our results for B0529+483: we detect structures at least 2--3 times smaller than the predicted broadening scale. We note that in any of the models given in Table~\ref{tab:tscat} the scattering at 22\,GHz is negligible at the scales of interest.

\subsection{Brightness temperature taking scattering into account}

It was shown by \citet{2016ApJ...820L..10J} that the brightness temperature estimate of \citet{Lobanov15} can be improved by taking into account scattering when it is present above the noise limit. To do this, the angular broadening, $\theta_\mathrm{scatt}$, must be known. We compute the maximum lower limit on brightness temperature at 1.7 and 4.8\,GHz using four options for scattering: 1) the NE2001 model, which seems to be ruled out for B0529+483; 2) the `Strong angular broadening' variant from Table \ref{tab:tscat}, which is inconsistent with the Kolmogorov spectrum; 3) the `Weak angular broadening' which agrees with our data; 4) no scattering, which also agrees with the data. The results are shown in Table~\ref{tab:tmax}.

\begin{table}
\caption{Maximum lower limit of the brightness temperature in units of $10^{13}$ K taking into account refractive scattering.}
\begin{center}
\begin{tabular}{lll}
\hline
Angular broadening      & 4.8 GHz & 1.7 GHz \\ \hline
NE2001 \citep{Cordes02} & 0.82    & 0.23 \\
Strong                  & 0.59    & 0.38 \\
Weak                    & 0.81    & 0.35 \\
No broadening           & 1.2     & 0.5 \\ \hline
\end{tabular}
\end{center}
\label{tab:tmax}
\end{table}

As is seen from Table~\ref{tab:tmax}, the difference between estimates of $T_{\rm b}$ is within a factor of 2. The brightness temperatures found in the limit of no scattering are very close to those estimated by fitting the model `Double Gaussian' in Table~\ref{tab:fit}.

\section{Discussion}

The data indicate the presence of at least two components in each of the 1.7\,GHz and 4.8\,GHz bands in the structure of B0529+483 on Space VLBI baselines, with their parameters given in Table~\ref{tab:fit}. We assume that the larger model components at two frequencies represent the same physical structure -- the quasar core. This is supported by the rough coincidence of the sizes and flux densities of the core measured by ground VLBI (Figs.~\ref{fig:var},\,\ref{fig:core}) and by \ra~(Table~\ref{tab:fit}). If this assumption is valid, the larger component of `Double Gaussian' model in Table~\ref{tab:fit} has a size scaling with frequency as $\nu^{-0.9}$. This is close to the intrinsic core property related to the synchrotron self-absorption expected from the \cite{BK79} model and measured for blazars \citep{2014MNRAS.437.3396K}. 
The size scaling $\propto \nu^{-1}$ is consistent with our ground VLBI data as well. For the small component the scaling cannot be determined accurately, and is $\propto \nu^{-1.2\pm2}$.

Since the quasar is located close to the Galactic plane, we might expect the presence of refractive scattering. We have considered the assumption that the large component is an angular broadened image of the source while the small component represents the refractive substructure. This hypothesis is ruled out by the scaling of the observed angular broadening in this case, which is strongly inconsistent with the expected scattering law. We note that \citet{2015MNRAS.452.4274P} analyzed a large set of extragalactic sources at a range of galactic latitudes. They found that core sizes of extragalactic radio sources far from the Galactic plane show a $\nu^{-1}$ scaling, while about 1/3 of the AGNs close to the Galactic plane exhibit scaling consistent with that expected from scattering. We also rule out the scattering predicted from the NE2001 model \citep{Cordes02}, since the large resolved components of B0529+483 have sizes smaller than predicted by the model.

We conclude that our results can be explained by either of the following two hypotheses. The first is that there are no refractive scattering effects in B0529+483 observations with \ra. This means that even close to the Galactic plane the resolution of tens of micro-arcseconds can be reached at relatively low frequencies of 1.7--5\,GHz. The second is that the minor axis of the large component at 1.7\,GHz is scatter-broadened. In this case the two detections at the longest baselines at 1.7\,GHz represent the refractive substructure. At 4.8\,GHz the angular broadening scale should be 0.04 mas. The assumption that the small component of the source at 4.8\,GHz is actually an angular broadened core is consistent with the data.

If the small component at 4.8\,GHz is affected by scattering, it also may be subject to interstellar scintillation. In that case the non-detections in Figs. \ref{fig:1dC}, \ref{fig:1dL} can be interpreted as the result of these scintillations: the component was detected only when it was scattered up. The amplitude and the brightness temperature of the small component may be decreased by a factor of 2 in that case. The small amplitude of this, $\sim0.02$\,Jy, in comparison with the large component, 0.7 Jy, unfortunately prevents the detection of these scintillations using single-dish intra-day variability observations.

In both these hypotheses the small component at 4.8\,GHz is not created by the refractive substructure. This means that the lower limit of the brightness temperature for B0529+483 at 4.8\,GHz is $1\times10^{13}$\,K (Table~\ref{tab:tmax}). This is consistent also with the estimate of the brightness temperature found at 22\,GHz, which is not affected by scattering, of $10^{13}$\,K, see Table~\ref{tab:fit}.
\cite{lister16} have measured the maximum apparent speed in the parsec-scale jet of this quasar to be $v_\mathrm{app}=19.8\pm3.0\,c$. In the case of equilibrium between the energy in particles and magnetic fields, the limit is $10^{10.5}$\,K, and so an unrealistically high Doppler boosting of $\delta>300$ is needed. The dominance of the particle energy to the magnetic energy in many AGNs is also confirmed by \citet{2017MNRAS.468.2372N}. 

On the other hand, following \cite{1994ApJ...426...51R}, we can estimate the electron energy loss timescale from the de-boosted values of brightness temperature and frequency of observation. If the Doppler boosting $\delta\approx1$, the timescales are far less than 1 second at both frequencies. Taking the Doppler boosting $\delta\approx v_\mathrm{app}/c$, we find the timescales to be $\sim100$ years at 4.8\,GHz and $\sim$1 year at 22\,GHz. The times are given in the observer's frame. Both these timescales are longer than the timescales of our observations of high brightness temperatures; thus, our results for B0529+483 do not challenge the inverse-Compton limit. 
The comparison of the radio and $\gamma$-ray photon flux in Fig.~\ref{fig:gamma} does not show any evidence of strong increase of $\gamma$-ray photon flux during the period of \ra\ fringe detections which could pump the energy of electrons and increase the brightness temperature above the inverse-Compton limit \citep[see, e.g.,][]{1994ApJ...426...51R,Kovalev16}.


\section{Summary}

\ra\ has detected the quasar B0529+483 at 1.7, 4.8, and 22~GHz on projected baselines up to 240\,000\,km. 
An analysis of the visibility amplitude versus projected baseline demonstrates the presence of at least two components, one of which is resolved and the other which is not.
We find that the data are consistent with two possibilities: either \ra\ detects no refractive scattering substructure for this low galactic latitude target, or the scattering is significant only at 18~cm. In any case, the scattering for this particular quasar is found to be much weaker than predicted by the NE2001 model \citep{Cordes02}.
%
The brightness temperature of B0529+483 core in the source frame is found to be greater than $10^{13}$~K at 4.8 and 22~GHz. This indicates a strong dominance of particle energy density over magnetic field energy density in quasar cores; otherwise extremely strong Doppler boosting, $\delta>300$, is needed. The inverse-Compton limit of $T_\mathrm{b}\sim10^{11.5}$~K requires $\delta\sim20-30$ which is not unreasonable.

\section*{Acknowledgements}

We thank Richard Porcas and Cristina Garcia Miro for useful discussions and comments.
The RadioAstron project is led by the Astro Space Center of the Lebedev Physical Institute of the Russian Academy of Sciences and the Lavochkin Scientific and Production Association under a contract with the Russian Federal Space Agency, in collaboration with partner organizations in Russia and other countries.
Results of optical positioning measurements of the {\it Spektr-R} spacecraft by the global MASTER Robotic Net \citep{2010AdAst2010E..30L}, ISON collaboration, and Kourovka observatory were used for spacecraft orbit determination in addition to mission facilities.
Partly based on observations performed with radio telescopes of IAA RAS (Federal State Budget Scientific Organization Institue of Applied Astronomy of Russian Academy of Sciences).
Partly based on the Evpatoria RT-70 radio telescope (Ukraine) observations carried out by the Institute of Radio Astronomy of the National Academy of Sciences of Ukraine in 2012 and 2013 under a contract with the State Space Agency of Ukraine and by the National Space Facilities Control and Test Center with technical support by Astro Space Center of Lebedev Physical Institute, Russian Academy of Sciences. 
Partly based on observations carried out using the 32-meter radio telescope operated by Torun Centre for Astronomy of Nicolaus Copernicus University in Torun (Poland) and supported by the Polish Ministry of Science and Higher Education SpUB grant.
Partly based on observations with the 100-m telescope of the MPIfR (Max-Planck-Institute for Radio Astronomy) at Effelsberg, observations with the Medicina and Noto telescopes operated by INAF --- Istituto di Radioastronomia, and observations performed with the RATAN-600 radio telescope of the Special Astrophysical Observatory of the Russian Academy of Sciences
(grant id.~73603 from \url{http://www.ckp-rf.ru}).
The Westerbork Synthesis Radio Telescope is operated by the ASTRON (Netherlands Institute for Radio Astronomy) with support from the Netherlands Foundation for Scientific Research (NWO). 
The authors are grateful to Antonis Polatidis (ASTRON) for assistance in obtaining and processing WSRT zero baseline data.
This research has made use of data from the MOJAVE database that is maintained by the MOJAVE team \citep{2009AJ....137.3718L}.
This research has made use of data from the OVRO 40-m monitoring program \citep{2011ApJS..194...29R} which is supported in part by NASA grants NNX08AW31G, NNX11A043G, and NNX14AQ89G and NSF grants AST-0808050 and AST-1109911.
This research was supported in part by the Basic Research Program P-7 of the Presidium of the Russian Academy of Sciences,
the government of the Russian Federation (agreement  05.Y09.21.0018), and the Alexander von Humboldt Foundation.

\bibliographystyle{mnras}
\bibliography{radio}

\end{document}